\documentclass[11pt]{article}
\usepackage{Blois,epsfig}

\def\be{\begin{equation}}
\def\ee{\end{equation}}
\def\bea{\begin{eqnarray}}
\def\eea{\end{eqnarray}}

\def\simge{\mathrel{%
   \rlap{\raise 0.511ex \hbox{$>$}}{\lower 0.511ex \hbox{$\sim$}}}}
\def\simle{\mathrel{
   \rlap{\raise 0.511ex \hbox{$<$}}{\lower 0.511ex \hbox{$\sim$}}}}

\begin{document}
\vspace*{2cm}

\vspace*{2cm}
\title{COLOR STRINGS, POMERONS AND COLOR GLASS CONDENSATE}

\author{ELENA FERREIRO }

\address{Departamento de F\'{\i}sica de Part\'{\i}culas,
Universidad de
Santiago de Compostela, Spain}  

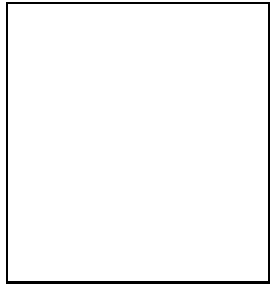
\maketitle\abstracts{
In the recent experiments like DIS at HERA or the heavy-ion experiments at RHIC, and
also in expected LHC at CERN, the number of involved partons is very large,
due to the high energy and/or the high number of participants of those experiments.
These high parton densities should in principal lead 
to an extremely huge multiparticle production, but experimentally we have seen that 
this is not the case. So there should be a mechanism that reduces the number of created
particle. 
Here, I review the problem of parton saturation and its implications through three in principal 
different
approaches, but somewhat related: saturation in a geometrical approach, QCD saturation through the
Color Glass Condensate and perturbative Pomeron approach with initial conditions.}

\section{Geometrical approach to saturation: String models and percolation}

In many models of hadronic and nuclear collisions, color strings are exchanged between 
the projectile and the target. Those strings act as {\it color sources} of particles through the creation
of $q-{\bar q}$ pairs from the sea. 
The number of strings grows with the energy and with the number of nucleons of the 
participant nuclei. 

In impact parameter space these strings are seen as circles inside the total collision area.
When the density of strings becomes high the string color fields begin to overlap and 
eventually individual strings may fuse, 
forming a new string --{\it cluster}-- which has a higher color charge at its ends, 
corresponding to the summation of the color charges located at the ends of the original strings. 
The new string clusters break into hadrons according to their higher color. 
As a result, there is a reduction of the 
total multiplicity. 
Also, as the energy-momenta of the original strings are summed to obtain the energy-momentum of 
the resulting cluster, the mean transverse momentum of the particles created by those clusters
is increased compared to the one of the particles 
created from individual sources.

As the number of strings increases, 
more strings overlap. 
Some years ago, it has been proposed in Ref. \cite{REF96} that
above a critical density of strings {\it percolation}
 occurs, so that paths of overlapping circles are formed through the whole collision area. 
Along these paths the medium behaves like a color conductor. 
Also in \cite{REF96}, we have made the remark that 
several fused strings can be considered as a domain of a {\it non thermalized Quark Gluon Plasma}. 
The percolation gives rise to the formation of a non thermalized 
Quark Gluon Plasma on a nuclear scale. 

Note that here we are not speaking about a final state interaction phenomenon, since there is no 
thermalization involved. In fact, what we are trying is
to determine under what conditions the initial state 
configurations can lead to color connection, and more specifically, 
if variations of the initial state can lead to a transition from 
disconnected to connected color clusters. 
The results of such a study of the pre-equilibrium state in nuclear collisions 
do not depend on the subsequent evolution and thus in particular not require any kind of 
thermalization.

The formalism underlying the transition from disconnected to connected systems
is given by percolation theory, which describes geometric critical behavior.

Consider placing $N$ small circular discs (color sources, strings or partons) 
of radius $r$ onto a large circular manifold (the transverse nuclear plane) of radius $R$; 
the small discs may overlap. 
With increasing density 
\be
\eta = \frac{N \pi r^2}{ \pi R^2}\  , 
\label{ec1}
\ee
this overlap will lead to 
more and larger connected clusters. 
The most interesting feature of this phenomenon is that the average cluster size 
increases very suddenly from very small to very large values. 
This suggests some kind of geometric critical behavior. In fact, 
the cluster size diverges at a critical threshold value $\eta_c$ of the density. This appearance
of an infinite cluster at $\eta= \eta_c$ is defined as percolation: the size of the cluster 
reaches the size of the system. 
$\eta_c$ has been computed using Monte Carlo simulation, direct connectedness expansion and 
other different methods. All the results are in the range  $\eta_c = 1.12 \div 1.175$.

In our model \cite{REF96} we had proposed a {\it fixed radius} for the independent 
color sources of 
$r=0.2 \div 0.25$ fm,
 that corresponds to a momentum around 1 GeV. 
This value has been obtained from 
Monte Carlo simulations in the framework of the String Fusion Model Code (SFMC) \cite{REFSFMC} 
made at SPS
energies. 
According to eq. (\ref{ec1}), in order to estimate the density $\eta$, one needs to know 
the number of sources $N$. In our model, it is obtained from the SFMC,
that, for nucleus-nucleus collisions, 
takes into account two contributions: one proportional to the number of participant
nucleons --valence-like contribution-- and another one proportional to the number of 
inelastic nucleon-nucleon collisions.
Note that $N$ will depend on the energy $\sqrt{s}$ (or equivalently, on $x$) and on the number of 
participant nucleons $A$, so in some way the condition to achieve percolation depends on 
$A$ and $s$, $\eta=\eta(A,x)$.
$\pi R^2$ corresponds simply to the nuclear overlap area, $S_A$, 
at the given impact parameter. This overlap area can be determined in 
a Glauber study, using Woods-Saxon nuclear profiles.
That leads to the following results: In our model,
at SPS energies, the critical threshold for percolation could eventually been 
achieved for the most central Pb-Pb collisions, and for sure in Au-Au central collisions at 
RHIC energies and even in p-p collisions at LHC energies.
Just as an example, in first approximation, one can estimates analytically that 
at very high energies, for central A-A collisions, if we take the number of initially created
sources as proportional to the number of collisions, 
$N \propto A^{4/3}$ and the nuclear overlap area $S_A= \pi R^2 \propto A^{2/3}$ 
then 
\be
\eta=\frac{N \pi r^2}{ \pi R^2}=
\frac{N_{sources}
S_1}{S_A} \propto A^{2/3}\ .
\label{ec2}
\ee

\section{The Color Glass Condensate}

Now we arrive to another approach, the QCD saturation through the formation 
of a Color Glass Condensate (CGC) \cite{REFCGC}. The idea is the following:
At high energy, the QCD cross-sections are controlled by small longitudinal
momentum gluons in the hadron wave function, whose density grows rapidly with
increasing energy or decreasing $x$, due to the enhancement of radiative
process. If one applies perturbation theory to this regime, one finds
that, by resumming dominant radiative corrections at high energy, the BFKL
equation leads to a gluon density that grows like a power of $s$ and in
consequence
to a cross-section that violates the Froissart bound.
Nevertheless,
 the use of perturbation theory to high-energy problems is not obvious. In
fact, the BFKL and DGLAP equations are linear equations that neglet the
interaction among the gluons. With increasing energy, recombination effects
--that are non-linear--
favored by the high density of gluons should become more important and lead
to an eventual {\it saturation} of parton densities.

These effects become important
when the interaction probability for the gluons becomes of order one.
Taking $\frac{\alpha_s N_c}{Q^2}$ as the transverse size of the gluon and
$\frac{x G(x,Q^2)}{\pi R^2}$ as the density of gluons, the interaction
probability is expressed by
\be
\frac{\alpha_s N_c}{Q^2}\,\,\times\,\,
\frac{x G(x,Q^2)}{\pi R^2}\ .
\label{ec5}
\ee
Equivalently, for a given energy, saturation
occurs for those gluons having a sufficiently large transverse size $r_\perp^2
\sim 1/Q^2$, larger than some critical value $1/Q_s(x,A)$. So the phenomenon
of saturation introduces a characteristic momentum scale,
the {\it saturation momentum} $Q_s(x,A)$, which is a measure of the
density of the saturated gluons, and grows rapidly with $1/x$ and
$A$ (the atomic number).
The probability of interaction
--that can be understood as "overlapping" of the gluons in the transverse space--
becomes of order one for those gluons with
momenta $Q^2 \simle Q_s(x,A)$ where
\be
Q^2_s(x,A)=\alpha_s N_c \ \frac{x G(x,Q^2_s)}{\pi R^2} \equiv
\frac{({\rm color\,\, charge})^2}{{\rm area}}\ .
\label{ec6}
\ee

For $Q^2\simle Q^2_s(x,A)$, the non-linear effects
are essential, since they
 are expected to soften the growth of the gluon distribution
with $\tau\equiv \ln(1/x)$.
For a nucleus, $x G_A(x,Q^2_s)\propto A$ and
 $\pi R^2_A\propto A^{2/3}$, so eq.~(\ref{ec6})
predicts
$Q^2_s\propto A^{1/3}$. One can estimate the saturation scale by
inserting the BFKL approximation into
eq.~(\ref{ec6}). This gives
(with $\delta\approx 1/3$ and $\lambda\approx c\bar\alpha_s$ in a
first
approximation):
\be
\label{ec7}
Q^2_s(x,A)\,\,\sim\,\,A^{\delta}\, x^{-\lambda}\,,
\ee
which indicates that an efficient way to create a high-density
environment is to combine large nuclei with moderately small values of $x$,
as it is done at RHIC. In fact the estimated momentum for saturation at RHIC
will be $Q_s= 1 \div 2$ GeV, in accordance with the result of the previous section.


\section{Perturbative QCD pomeron with saturation in the initial conditions}

Let us now try a different approach \cite{REFMIJAILPAJ}. 
Consider now the nucleus-nucleus interaction as
governed by the exchange of pomerons.
Its propagation is  governed
by the BFKL equation. 
Its interaction is
realized by the triple
pomeron vertex. Equations
which describe nucleus-nucleus interaction in the perturbative QCD
framework have been obtained in ~\cite{REFBRA2}.
Knowing that the AGK rules are satisfied
for the diagrams with BFKL pomerons interacting via the triple
pomeron vertex
one can conclude that
 the inclusive cross-section will 
be given by the convolution of two sums of fan diagrams propagating
from the emitted particle towards the two nuclei. 

Taking $A=B$ and constant nuclear density for $|b|<R_A$, one can find
the inclusive cross-section in perturbative QCD as 
\be
I_{A}(y,k)=A^{2/3}\pi R_0^2\frac{8N_c\alpha_s}{k^2}\int
d^2re^{ikr}[\Delta\Phi_A(Y-y,r)]
[\Delta\Phi_A(y,r)],
\label{ec1p}
\ee
where $\Delta$ is the two-dimensional Laplacian and $\Phi(y,r)$ is the
sum
of all fan diagrams connecting the
pomeron at rapidity $y$ and of the transverse dimension $r$ with the
colliding nuclei, one at rest and the other at rapidity $Y$.
The function $\phi_A(y,r)=\Phi(y,r)/(2\pi r^2)$, in the momentum space,
 satisfies the well-known
non-linear Balitsky-Kovchegov equation
\be
\frac{\partial\phi(y,q)}{\partial \bar{y}}=-H\phi(y,q)-\phi^2 (y,q),
\ee
where $\bar{y}=\bar{\alpha}y$, $\bar{\alpha}=\alpha_sN_c/\pi$,
$\alpha_s$ and $N_c$ are the strong coupling constant and the number
of colors,
respectively, and $H$ is the BFKL Hamiltonian. This equation has to be solved
with the initial condition at $y=0$ determined by the color dipole
distribution in the nucleon smeared by the profile function of the
nucleus.

We can take the initial condition in accordance with
the Golec-Biernat distribution \cite{REFGBW}, that takes into account
saturation according to the CGC.
Then we can find the following result:
One can observe that whereas at relatively small momenta the inclusive
cross-sections are proportional to $A$, that is to {\it the number of
participants}, at
larger momenta they grow with $A$ faster, however  noticeably slowlier
than the number of collisions, approximately as $A^{1.1}$.
The interval of momenta for which $I_A\propto A$ is growing with energy,
so that one may conjecture that at infinite energies all the spectrum
will be proportional to $A$.

Note that in the string models, like Quark Gluon String Model or Dual Parton Model \cite{REFDPM}
each string corresponds to the exchange of pomerons, and the interaction among the strings 
would correspond to interaction among the pomerons through the triple pomeron vertex.

\section{Conclusions}

We have compared different models that takes into account saturation in different ways:
from the semi-phenomenological
fusing color sources picture for the soft domain including percolation, 
the QCD saturation through the Color Glass Condensate
and those which
follow from the pomeron approach, perturbatively
derived from QCD, taking into account saturation in the initial conditions.
In fact, it seems that the exchanged of
elemental objects, color sources --strings, partons or pomerons--, should lead to a saturation
in the initial conditions when the densities are high enough.

\section*{Acknowledgments}
I would like to thank Maurice Haguenauer, Basarab Nicolescu and Elizabeth Hautefeuille
for organizing such a nice conference.

\end{document}